# Superconductivity in layered iron Selenide induced by cobalt- and sodium-doping

Zhanqiang Liu[1], Aihua Fang[1], Fuqiang Huang[1*], Mianheng Jiang[2]

[1]*State Key Laboratory of High Performance Ceramics and Superfine Microstructure, Shanghai Institute of Ceramics, Chinese Academy of Sciences, Shanghai 200050, P. R. China*

[2]*Shanghai Institute of Microsystem and Information Technology, Chinese Academy of Sciences, Shanghai 200050, P. R. China*

**Abstract:**

Superconductivity with zero resistance transition temperature ($T_c$) up to 8.4 K and 8.3 K can be obtained by doping cobalt and sodium in α-FeSe with the nominal composition of $Fe_{0.92}Co_{0.08}Se$ and $Na_{0.1}FeSe$, respectively. The electrical resistivity and AC magnetic susceptibility of the prepared samples, measured with physical property measurement system (PPMS), unambiguously consistent with each other to indicate that the samples are superconductive. The respective doping mechanisms for cobalt and sodium into the parent α-FeSe are the Fe-site substitution and the interlayer insertion. It is the first time that α-FeSe can be induced to be a superconductor with $Na^+$ intercalated into the interlayers.

The pnictides containing the anti-PbO-like [FeAs] layered motifs have attracted much attention, since the superconductive transition temperature of 26 - 43 K was found in LaFe($O_{1-x}F_x$)As [1, 2]. The other series of $M$Fe$_2$As$_2$ ($M$ = Ba, Sr and Ca) superconductors were also reported [3-8]. These materials show the obvious superconductivity by the aid of chemical doping in the positions of La, Fe, O and $M$, such as the Sr-doped LaFeOAs [9-10], F-doped LaFeOAs [1] and K-doped BaFe$_2$As$_2$ [3]. It is very interesting that Co is the only the $d$-block transition element for doping in the Fe position to induce the superconductivity in the literatures [11-17].

α-FeSe, with the same space group P4/nmms and similar planar crystal sublattice just like LaFeOAs, has no superconducting. The very recent work showed that preparing α-FeSe with intentional Se deficiency is one effective method to achieve it's superconducting with $T_c$ equal to 8 K [18]. Then the question is whether the partial substitution of Fe by Co in α-FeSe can induce the superconductivity except the Se deficiency? Furthermore, how about the sodium intercalation into the weakly-coupled FeSe layers to form Na$_x$FeSe, similar to the superconductors Na$_x$TaS$_2$ [19] and Cu$_x$TiSe$_2$ [20-22]? These questions initialize the motivation of the present work, and the superconductivity with $T_c$ up to 8.4 K and 8.3 K observed by Co-doping or Na-insertion in the α-FeSe structure. The doping mechanisms for cobalt and sodium are also discussed in the context.

Polycrystalline samples with Fe$_{1-x}$Co$_x$Se ($x$ = 0.08) and Na$_y$FeSe ($y$ = 0.1) nominal compositions were prepared by conventional solid state reaction. The well-mixed stoichiometrically fine powders of Fe (99%), Co (99%), Se (99.99 %) and Fe, Se, Na$_2$Se (self-made) were cold-pressed into pellets, respectively. Then sealed in evacuated quartz tubes, placed in furnace and heated at the temperature of 750 $^o$C twice with cooled down automatically and lastly annealed at the temperature of 400 $^o$C for 30 hours. These samples were kept in a purified argon atmosphere glove box prior to each characterization.

X-ray diffraction (XRD) analysis was carried out to identify the samples, using an X-ray diffractometer (Rigaku D/Max 2550V, 40 kV 40 mA) with Cu$K\alpha$ radiation in the 2θ range from 20° to 70°.

The electrical resistivity was measured by the standard four-probe method with silver-paint contacts. The AC magnetic susceptibility was measured with a modulation field of 1 Oe and 333 Hz. And both of these measurements were carried out from 5 – 300 K on a Quantum Design physical property measurement system (PPMS, Model 6000).

The phases of polycrystalline samples $Fe_{0.92}Co_{0.08}Se$ and $Na_{0.1}FeSe$ were examined by X-ray powder diffraction patterns, as shown in Fig. 1. The XRD patterns indicate that α-FeSe (JCPDS no. 03-0533) was successfully synthesized with small amount of impurities, and the major impurity was identified to be element selenium, which are labeled as stars in the figure. The amount of impurity selenium in $Na_{0.1}FeSe$ seems to be higher than that in $Fe_{0.92}Co_{0.08}Se$. The reason is that the home-made raw material $Na_2Se$ used in the preparation of $Na_{0.1}FeSe$ containing some extra Se. Though the doping amounts of Co and Na are as high as the designed 8 % and 10 %, respectively, the peak shift difference between $Fe_{0.92}Co_{0.08}Se$ and $Na_{0.1}FeSe$ is not very obvious.

Fig. 2 shows the temperature dependence of the electrical resistivity ($\rho$) of polycrystalline samples $Fe_{0.92}Co_{0.08}Se$ and $Na_{0.1}FeSe$ in zero magnetic field. The inset of the figure is the enlarged low temperature data. The superconducting transitions can be seen from the inset, and the $T_c$ temperatures of $Fe_{0.92}Co_{0.08}Se$ and $Na_{0.1}FeSe$ are 8.4 K and 8.3 K, higher than that of the reported Se deficient FeSe sample. The plot of $Fe_{0.92}Co_{0.08}Se$ has a temperature-linear dependent resistivity observed from 300 K to 10 K. However, there is a small anomaly in $Na_{0.1}FeSe$ for increase in resistivity at around 250 K.

In order to confirm the superconductivity, the AC magnetic susceptibility of the sample $Fe_{0.92}Co_{0.08}Se$ was measured with the field amplitude of 1 Oe and the frequency of 333 Hz. The plot of the temperature dependence of AC magnetic susceptibility is shown in Fig. 3. It is obvious to see that there is a sudden abrupt drop at 8.4 K, which is coincident with the measurement of electrical resistivity in Fig. 2. The smooth line without any trace of background and the sudden abrupt drop indicate the superconductivity character of $Fe_{0.92}Co_{0.08}Se$, and the superconductivity is not apparently affected by the impurities as exhibited in the XRD patterns. The similar result, which is not shown here, was also found for the sample $Na_{0.1}FeSe$.

It is very necessary to firstly discuss here that, with an appreciate amount of $Co^{2+}$ doping in the $Fe^{2+}$ sites can be induced to be a superconductor without a big surprise, which was reported in the Co-doped compounds of LaFeOAs [15, 16], $BaFe_2As_2$ [14] and $SrFe_2As_2$ [13]. The superconductivities were achieved by introducing electron dope and the stronger effects of disorder with replacing Fe with Co. whereas there is only disorder effect produced by doping Co in the layered α-FeSe. Actually, Cu and Mn etc. were also used to replace Fe, but no superconductive phenomenon was found. Secondly, the Na-doping in α-FeSe with the $Na_{0.1}FeSe$ nominal composition is very different from that of Co. $Na^+$ has the ability to be located in the interlayer of the layered [FeSe] structure. So the $Na^+$ with appropriate molar ration, such as 0.1, can be intercalated into the weak coupled region between the anti-PbO-like [FeSe] layers to form $Na_xFeSe$. The similar results were reported in some layered transition-metal dichalcogenides, which, for example $TaS_2$ [19] and $TiSe_2$ [20-22], can be induced to be superconductors with specific ions ($Na^+$ or $Cu^+$) intercalation. So it can be inferred from $Na_{0.1}FeSe$ that layered transition-metal chalcogenides are also potential superconductors.

α-FeSe can be successfully induced to a superconductor with appropriate amount of cobalt doping or sodium doping. The two doped samples, $Fe_{0.92}Co_{0.08}Se$ and $Na_{0.1}FeSe$, have the similar zero resistance transition temperature 8.4 K and 8.3 K. The results of electrical resistivity and AC magnetic susceptibility are consistent with each other. The doping mechanisms for cobalt and sodium into parent α-FeSe are different and sodium doping follows the manner of $Na^+$ intercalation into the interlayers of FeSe. $Na_{0.1}FeSe$ sends a message that layered transition-metal chalcogenides are potential superconductors.

**Acknowledgements:**

This research was financially supported by National 973 Program of China Grant 2007CB936704 and National Science Foundation of China Grant 50772123.

**Corresponding author:**

Fuqiang Huang: huangfq@mail.sic.ac.cn

Tel.: +86 21 52411620; fax: +86 21 52413903

**Figure captions:**

Fig. 1: XRD patterns of samples $Fe_{0.92}Co_{0.08}Se$ (A) and $Na_{0.1}FeSe$ (B).

Fig. 2: Temperature dependence of electrical resistivity of samples $Fe_{0.92}Co_{0.08}Se$ (A) and $Na_{0.1}FeSe$ (B).

Fig.3: Temperature dependence of AC magnetic susceptibility of samples $Fe_{0.92}Co_{0.08}Se$.

**Figure 1:**

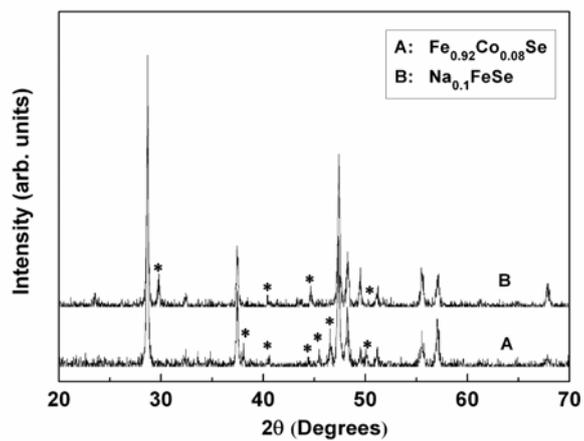

**Figure 2:**

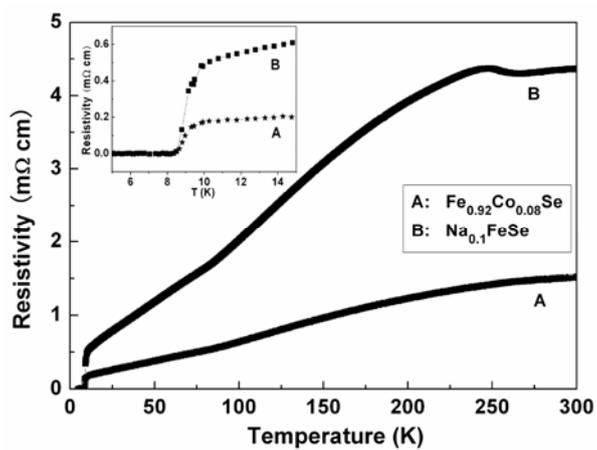

**Figure 3:**

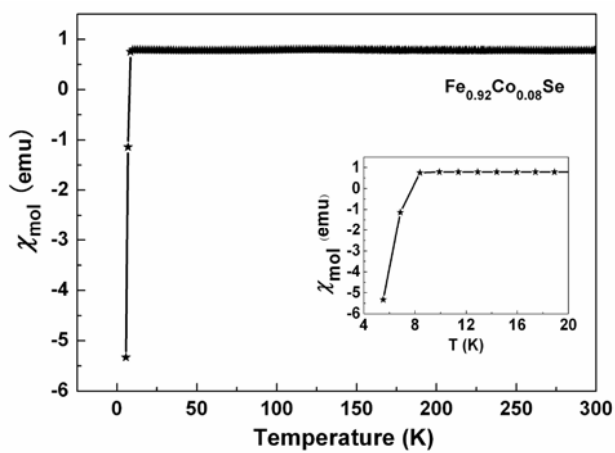